\documentclass[10pt, twocolumn, floatfix, superscriptaddress, aps, prb]{revtex4-2}
\usepackage{amsmath, amssymb, graphicx, hyperref, xcolor, microtype}

\newcommand{\nospace}{}
\newcommand{\changes}[1]{#1}

\begin{document}

\title{Disordered monitored free fermions}

\author{Marcin Szyniszewski}
\affiliation{Department of Physics and Astronomy, University College London, Gower Street,
London, WC1E 6BT, UK}

\author{Oliver Lunt}
\affiliation{Department of Physics, King's College London, Strand, London, WC2R 2LS, UK}
\affiliation{School of Physics \& Astronomy, University of Birmingham, Birmingham, B15 2TT, UK}
\affiliation{Department of Physics and Astronomy, University College London, Gower Street,
London, WC1E 6BT, UK}

\author{Arijeet Pal}
\affiliation{Department of Physics and Astronomy, University College London, Gower Street,
London, WC1E 6BT, UK}

\begin{abstract}
Scrambling of quantum information in unitary evolution can be hindered due to
measurements and localization, which pin quantum mechanical wavefunctions in
real space suppressing entanglement in the steady state. In monitored
free-fermionic models, the steady state undergoes an entanglement transition
from a logarithmically entangled critical state to area-law. However, disorder
can lead to Anderson localization. We investigate free fermions in a random
potential with continuous monitoring, which enables us to probe the interplay
between measurement-induced and localized phases. We show that the critical
phase is stable up to a finite disorder and the criticality is consistent with
the Berezinskii-Kosterlitz-Thouless universality. Furthermore, monitoring
destroys localization, and the area-law phase at weak dissipation exhibits
power-law decay of single-particle wave functions. Our work opens the avenue to
probe this novel phase transition in electronic systems of quantum dot arrays
and nanowires, and allow quantum control of entangled states.
\end{abstract}

{\maketitle}

\section{Introduction}

The preservation of information in many-body quantum systems poses a substantial
 challenge in quantum computing. Generically, as quantum systems evolve
in time, any initial quantum information 
is scrambled throughout the system becoming inaccessible through local measurements, leading to thermalization. 
In recent years it has become clear that there are quantum systems
that can fail to thermalize, the most prominent example being the phenomenon of
many-body localization (MBL)~\cite{Basko2006, Gornyi2005, Pal2010,
Nandkishore2015, Abanin2019}. In such systems, quantum information remains
accessible via local measurements even at long times and preserves correlations in the initial state. The MBL phase transition separating localized and chaotic phases of matter, is characterised by a singular change in the entanglement properties of the system.  

Entanglement phase transitions can also occur in quantum trajectories of open
quantum systems~\cite{Li2018, Chan2018, Skinner2018}. In particular, the
transition occurs due to a competition between measurements and unitary
evolution, hence the name \emph{measurement-induced entanglement transition}
(MIET). This novel type of phase transition has been of interest in many recent
studies~\cite{Li2019, Szyniszewski2019, Zabalo2020, Napp2019, Fan2021,
Gullans2019purification, Bera2020, Vasseur2019, Bao2020, Jian2020, Zabalo2021,
Sierant2022fract, Li2020Conformal, Szyniszewski2020universality,
LopezPiqueres2020, Shtanko2020, Lavasani2021, Sang2021measurement, Zhang2020,
Choi2020, Turkeshi2020, Gullans2020, Nahum2021, Cao2019, Alberton2021,
Buchhold2021, Jian2020tn, Zhang2021, Botzung2021, Turkeshi2021zeroclicks,
Kells2021, Turkeshi2022, Turkeshi2022negativity, Popperl2022,
Ippoliti2021entanglement, Bao2021, Tang2020, Goto2020, Fuji2020, Rossini2020,
Lunt2020, Chen2020, Liu2021, Biella2020, Gopalakrishnan2021, Jian2021, Tang2021,
Turkeshi2021, Lang2020, VanRegemortel2021, Vijay2020, Nahum2020defects, Li2021,
Gullans2020, luntMeasurementinducedCriticalityEntanglement2021,
Sierant2022cliff, Gullans2020lowdepth, Fidkowski2021, Maimbourg2021,
Iaconis2020, Ippoliti2021postselection, Lavasani2020topological,
Sang2020entanglement, Shi2020, Rossini2021, Lu2021, Ippoliti2021, Zhang2021syk,
Jian2021syk, Bentsen2021, Minato2021, Doggen2021, Sharma2022, Block2021,
Muller2021, Czischek2021, Noel2021, Sierant2021, Medina2021, Agrawal2021,
Yang2021, Boorman2021, Levy2021, Minoguchi2022, Zhu2022, Piccitto2022, Cote2022,
Altland2022, Fleckenstein2022, Klocke2022, Ippoliti2022, Carollo2022,
Barratt2022, Barratt2022learn, Sriram2022, Dias2022, Milekhin2022, Feng2022,
Kelly2022, Wampler2022, Sahu2021, Yoshida2021, Coppola2022, Hashizume2022,
Li2021tn, Yu2022, Kalsi2022, Ladewig2022, Han2022, Zhang2022, Koh2022,
Schomerus2022, Jin2022, Zhou2022, Iadecola2022, Fisher2022, Buchhold2022,
Li2022, Kuno2022, Gal2022}. Typically, one considers a quantum circuit with
unitary gates interspersed with  local measurements at random locations. The
transition between the volume-law and the area-law phase occurs at a finite
measurement probability, and is known to occur in a wide variety of systems:
unitaries can be randomly drawn either from the Haar measure or the Clifford
group~\cite{Li2018, Chan2018, Li2019, Skinner2018}, or a Hamiltonian evolution
of interacting systems~\cite{Tang2020, Goto2020, Fuji2020, Rossini2020,
Lunt2020}, while the measurements can be chosen to be projective or weak. The
universal properties of the MIET in random unitary circuits have similarities
with those of percolation, though there appear to be some differences in surface
critical behavior~\cite{Skinner2018, Bao2020, Li2020Conformal, Jian2020,
Vasseur2019, Zabalo2020, Zabalo2021}.

\begin{figure}[b]
  \centering
  \includegraphics[width = 0.99\columnwidth]{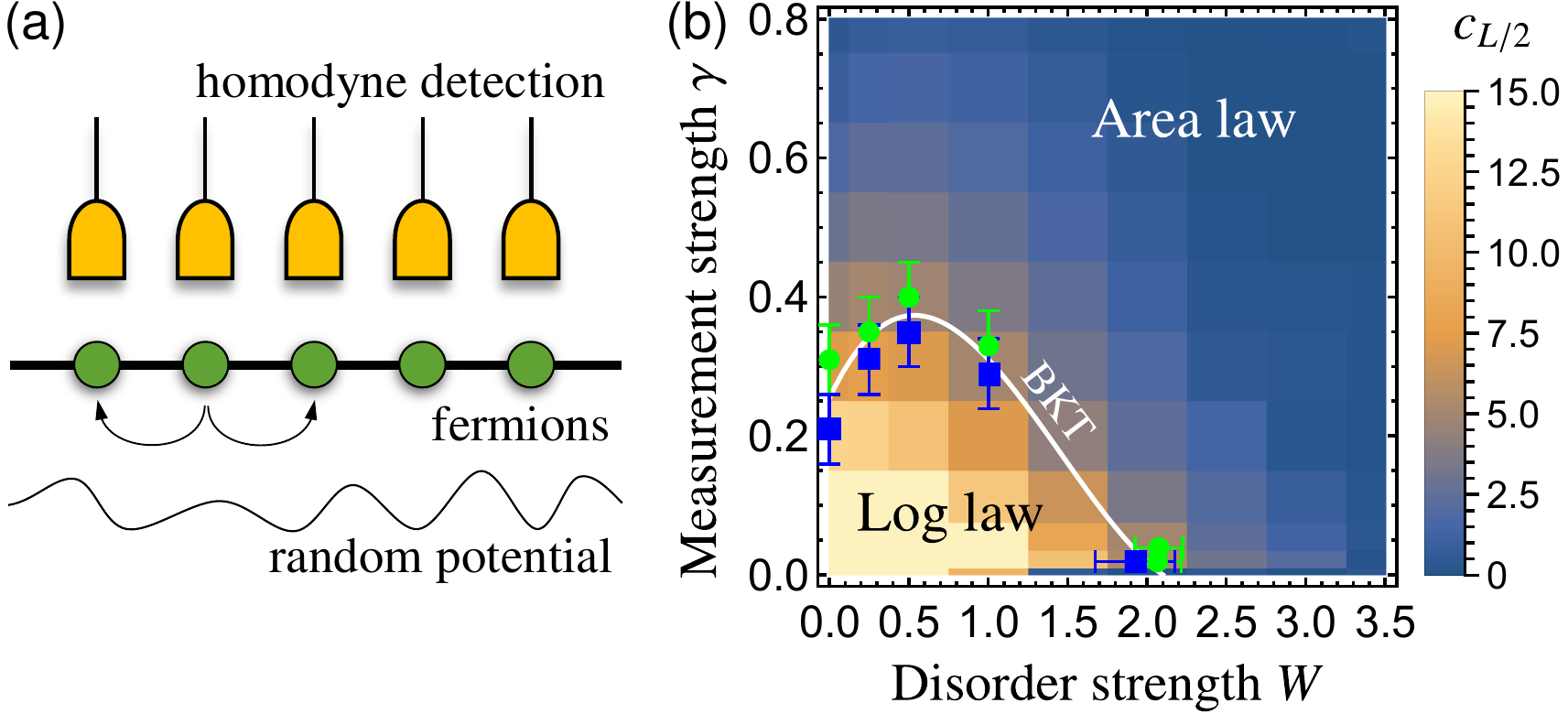}
  
  \caption{(a)~Sketch of disordered monitored free fermions. 
    (b)~Phase diagram. The density plot shows the effective central charge estimate, $c_{L/2}$. Data collapses of half-chain entanglement entropy (green circles) and central charge (blue squares) are used to estimate the  transition boundary (solid line).} \label{fig:phase_diagram}
\end{figure}

Intriguingly, the phase diagram changes significantly for a free-fermionic
system~\cite{Cao2019, Alberton2021, Buchhold2021, Bao2021, Jian2020tn,
Zhang2021, Botzung2021, Ippoliti2021entanglement, Turkeshi2021zeroclicks,
Kells2021, Turkeshi2022, Turkeshi2022negativity, Popperl2022}. The volume-law
entangled steady state for non-zero measurement probability is fragile due to
its lack of multipartite entanglement in free-fermion systems~\cite{Cao2019,
Alberton2021}. For a range of measurement probabilities, an extended critical
phase with logarithmic growth of entanglement and conformal symmetry
emerges~\cite{Alberton2021}. Beyond a critical measurement probability the
systems transition into an area-law phase. A substantial amount of evidence
implies that this MIET is within the Berezinskii-Kosterlitz-Thouless (BKT)
universality class~\cite{Alberton2021, Buchhold2021}, which puts it in a
distinct class from random unitary circuits. Recent developments also suggest
that the transition happens due to pinning of the wave-function trajectory to
the eigenstates of the measurement operator~\cite{Buchhold2021}.

However, several important questions relating to the robustness of the critical
logarithmic phase remain unanswered. For example, the logarithmic phase
remaining stable against breaking of the continuous $U(1)$ symmetry, for
particle number conservation, to a discrete $\mathbb Z_2$ fermion parity
symmetry is associated with continuous replica symmetry breaking which doesn't
appear to have a physical analog~\cite{Nahum2020defects, Nahum2021,
Buchhold2021, Turkeshi2021zeroclicks}. For free-fermionic systems in one
dimension, it is particularly interesting to ask about robustness to quenched
disorder. For a non-interacting Hamiltonian, arbitrarily weak disorder localizes
the single-particle modes in 1D, a phenomenon known as Anderson
localization~\cite{Anderson1958, Thouless1972, Abrahams1979}. The role of
measurements can destroy the localized phase at intermediate couplings while
facilitating localization into product states at strong coupling. The
competition between measurements and quenched disorder can result in a rich
phase structure for an entanglement transition and is also relevant for
observing the critical to area-law phase transition in an experimental setting.
Disorder plays an essential role in a system of electrons in quantum dot arrays
and nanowires where this phenomena can be explored.

Motivated by this question, in this article we investigate the impact of
quenched disorder on the measurement-induced transition in a one dimensional
free-fermion system. A careful analysis of the entanglement entropy and central
charge leads us to a phase diagram in terms of measurement strength $\gamma$ and
disorder $W$ [see Fig.~\ref{fig:phase_diagram}(b)] which exemplifies the
robustness of the logarithmic phase and the relationship between Anderson
localized and measurement induced area-law phases of non-interacting electrons.

\section{Model}

We consider spinless fermions hopping in a one-dimensional lattice with a random potential, subject to continuous
measurements [see Fig.~\ref{fig:phase_diagram}(a)],
\begin{equation}
  H = \sum_{i = 1}^L (c_i^{\dagger} c_{i + 1} + \text{h.c.})
    + \sum_{i = 1}^L h_i n_i,
  \label{eq:anderson}
\end{equation}
where the random potential is distributed uniformly $h_i \in [-W, W]$ with
disorder strength $W$. The system is initially set to a separable N\'eel state.
The evolution is implemented using the stochastic Schr\"odinger equation,
\begin{eqnarray}
  d|\psi(t)\rangle &=& -i\, H\, dt|\psi(t)\rangle
  - \frac{\gamma\, dt}{2} \sum_i (n_i-\langle n_i\rangle)^2 |\psi(t)\rangle \nonumber\\
  & & +\sum_i (n_i-\langle n_i\rangle) d\eta_i^t|\psi(t)\rangle,
  \label{eq:stoch}
\end{eqnarray}
which describes the continuous monitoring of particle number operator $n_i$ on
each site, with measurement strength $\gamma$~\cite{Cao2019}. The It\^{o} 
increments $d\eta_i^t$ have zero mean and variance of
${\gamma\,dt}$ (see Appendix~\ref{app:method} for details).

We monitor each quantum trajectory, characterized by a set of measurement
outcomes for a single realization of the random potential; the results are then
averaged over multiple trajectories. Importantly, this provides access to
averages of non-linear functions of the reduced density matrix, which in turn
allow us to capture the entanglement phase transition. Specifically, we use the
von Neumann entropy, a measure of entanglement between subsystem A and its
complement, defined as $S = - \text{tr}(\rho_\text{A} \ln \rho_\text{A})$,
where $\rho_\text{A}$ is the reduced density matrix of A. $S$ is initially zero
for a separable state, and grows in time, saturating near a fixed point
$S_\infty$ at long times, estimated as time average after saturation, $S_\infty
= \lim_{\Delta T \to \infty} \int_{t_\text{sat}}^{t_\text{sat} + \Delta T} S(t)
dt/\Delta T$. Finally, $S_\infty$ is averaged over trajectories, giving $\bar
S$.

Entanglement phase transitions can be directly observed by monitoring how $\bar
S$ changes with the system size $L$. However, even in free fermion circuits,
where we can access larger system sizes, finite size effects are significant and
impede our analysis. Special care needs to be taken for the critical phase,
where both $\bar S$ and the correlation length $\xi$ diverge logarithmically
with $L$ --- extraction of the critical point is difficult for phase
transitions with slowly diverging length scales~\cite{Alberton2021}. This
critical phase is expected to be described by a 1+1D non-unitary conformal field
theory (CFT) with periodic boundaries, with
\begin{equation}
  \bar S (l, L) = \frac{c}{3}
  \ln \left( \frac{L}{\pi} \sin \frac{\pi l}{L} \right) + s_0,
  \label{eq:central_charge}
\end{equation}
where $l$ is the length of the subsystem A, $c$ is the effective central charge
of the non-unitary CFT, and $s_0$ is the residual entropy. For large enough
systems, $c$ is expected to be zero in the area law phase and finite in the log
phase, and thus can be used as a transition diagnostic.

\section{Results}

\subsection{Phase diagram}

Using the results for $\bar S(L/2,L)$ as a function of $L$, we perform a fit to
Eq.~\ref{eq:central_charge} and obtain a central charge estimate, $c_{L/2}$.
This allows us to draw the dependence of $c_{L/2}$ on the measurement strength
$\gamma$ and the disordered field strength $W$ -- see
Fig.~\ref{fig:phase_diagram}(b). The central charge remains non-zero at low
values of $\gamma$ and $W$, implying the existence of the critical phase.
However, at large values of either $\gamma$ or $W$, $c_{L/2}$ stays close to
zero, a signature of the area law. This suggests that the logarithmic
phase survives the introduction of the random disordered field, and only when
the field is strong enough ($W \gtrsim 3.5$), the phase breaks down.

\begin{figure}[t]
  \centering
  \includegraphics[width = 0.99\columnwidth]{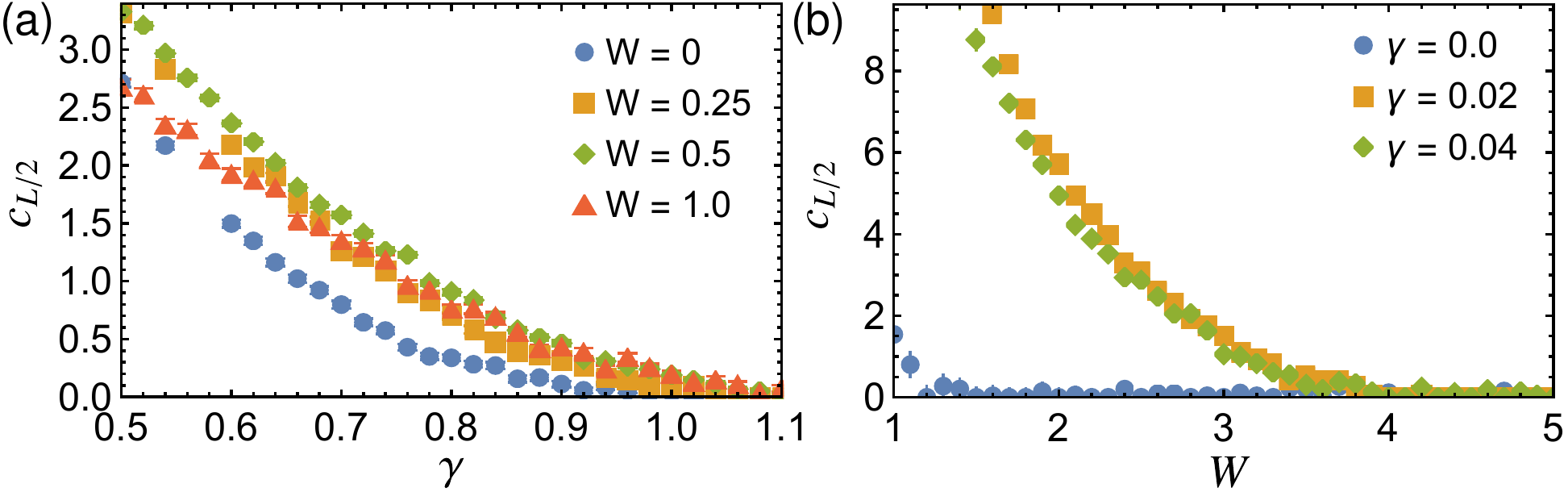}
  
  \caption{Effective central charge estimate, $c_{L/2}$, calculated using
  half-chain entanglement entropy (a)~for small values of $W$, and (b)~for small values of $\gamma$.}
  \label{fig:limit_behavior}
\end{figure}

Estimation of the precise phase boundary is, however, difficult, as $c_{L/2}$
does not decay sharply to zero. Large finite size effects necessitate a scaling
analysis, which we perform in the next sections. Nonetheless, two peculiar
features can immediately be seen in the density plot in
Fig.~\ref{fig:phase_diagram}(b): for small $W$, there is a non-monotonic
behavior of the phase boundary [see Fig.~\ref{fig:limit_behavior}(a)]; and for
small $\gamma$, the density plot shows a rapid change of $c_{L/2}$ for
$\gamma=0$ (Anderson localized) versus when $\gamma$ is finite [see
Fig.~\ref{fig:limit_behavior}(b)].

\subsection{Survival of the BKT universality class}

We firstly discuss the case of small disorder strength. For the clean system
($W=0$), we can fully reproduce existing results~\cite{Cao2019, Alberton2021}.
Importantly, Ref.~\cite{Alberton2021} provides numerical evidence of the BKT
universality class, for which the half-chain entropy can be collapsed
using~\cite{Harada1997},
\begin{equation}
  \bar S(L/2,L,\gamma) - \bar S(L/2,L,\gamma_c^{\bar S})
  = F[ (\gamma - \gamma_c^{\bar S}) (\ln L)^2 ],
  \label{eq:S_data_collapse}
\end{equation}
where $\gamma_c^{\bar S}$ is the critical point; the optimal collapse gives
$\gamma_c^{\bar S} \approx 0.31$. Another estimate comes from investigating $c$
as a function of system size $L$. To do this, one extracts $c(L)$ for one
specific $L$ by fitting the entropy results for different bipartitions to
Eq.~\ref{eq:central_charge}. Then, the $c(L)$ data can be collapsed according
to~\cite{Alberton2021},
\begin{equation}
  c(L) \gamma g(L)
  = \tilde F[ \ln L - \alpha / \sqrt{\gamma - \gamma_c^{c(L)}} ],
  \label{eq:c_data_collapse}
\end{equation}
which yields $\gamma_c^{c(L)} \approx 0.21$. \changes{The scaling function $g(L) = [1+1/(2\ln L - \beta)]^{-1}$~\cite{Harada1997, Carrasquilla2012, Alberton2021, Note1}.} Although the two
estimates have substantial error bars ($\approx 0.05$), both are much lower than
estimate based on $c_{L/2}$, where the transition would rather be expected at
$\gamma_c^{c_{L/2}} \approx 0.8$ [cf.~Fig.~\ref{fig:limit_behavior}(a)]. This
strongly suggests that $c_{L/2}$ cannot provide a good estimate of the
transition point due to finite size effects, and data collapses of $\bar S$ and
$c(L)$ are needed instead.

\begin{figure}[tb]
  \centering
  \includegraphics[width = 0.99\columnwidth]{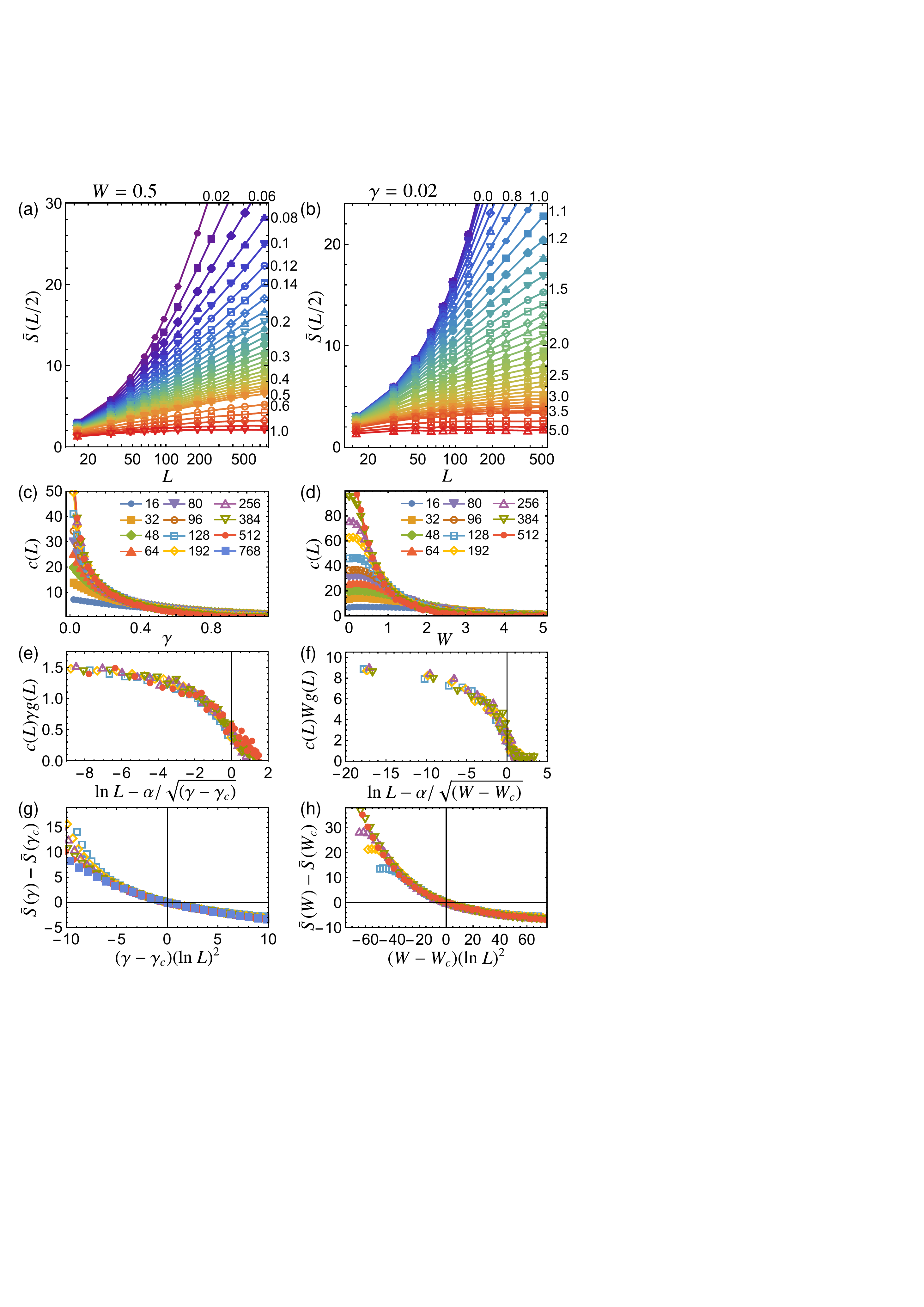}
  
  \caption{Behavior for small $W$ (left plots) and small $\gamma$ (right plots). (a,b)~Half-chain entropy $\bar S(L/2)$ for different values of the measurement strength $\gamma$ or disorder strength $W$ (see labels on the right). (c,d)~Central charge $c(L)$ as a function of $\gamma$ (or $W$) and system size $L$. Data collapses for (e,f)~$c(L)$, and (g,h)~$S(L/2)$. Legend from (c,d) applies in (e-h).}
\label{fig:small_param}
\end{figure}

Armed with this knowledge, we introduce a small amount of disorder ($W = 0.25,
0.5, 1.0$)~\footnote{Supporting data for $\gamma=0.04$ and $W \in \{0.25,
1.0\}$, as well as details of the finite-size scaling analysis, and the
discussion of single-particle wave functions can be found in Appendix~\ref{app:method}.}. The
results for the half-chain entropy and the central charge are shown in
Figs.~\ref{fig:small_param}(a,c). Judging from $\bar S(L/2)$, the corresponding
transition region where the entropy starts deviating significantly from a
$\ln(L)$ behavior for large $L$, is $0.3 \lesssim \gamma_c \lesssim 0.37$ for
$W = 0.5$. We also perform the data collapse for $\bar S(L/2)$ and $c(L)$, shown
in Figs.~\ref{fig:small_param}(e,g), finding $\gamma_c^{\bar S} \approx 0.40$
and $\gamma_c^{c(L)} \approx 0.35$ for $W = 0.5$. All performed data collapses
are of reasonably good quality, which suggests that the BKT universality class
of the transition is preserved in the presence of weak disorder. This is
consistent with the idea that the relevant symmetry is the continuous replica
symmetry, which should be preserved as long as the system is still
free-fermionic~\cite{Nahum2020defects, Nahum2021, Buchhold2021}.

\begin{figure}[tb]
  \centering
  \includegraphics[width = \columnwidth]{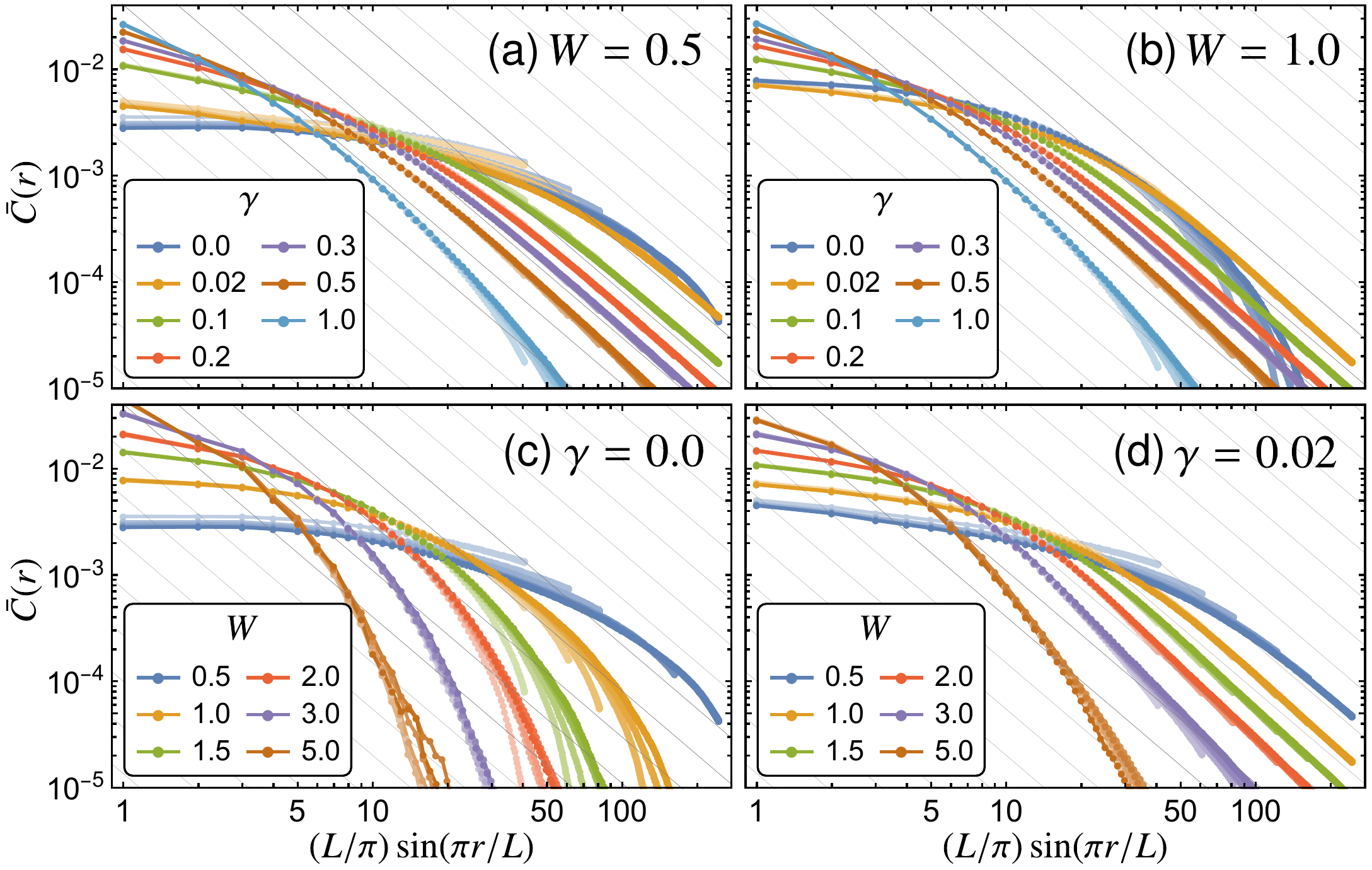}
  
  \caption{Connected correlation function $\bar C(r)$ for constant disorder strength (a) $W=0.5$, (b) $W=1.0$, and constant measurement strength (c) $\gamma=0.0$, (d) $\gamma=0.02$. Plot opacity indicates the system size ($L = 128, 196, 256, 384, 512, 768$). Gray lines show the algebraic decay of $\sim r^{-2}$ expected for the critical phase.} \label{fig:correlations}
\end{figure}

This result is corroborated further by the behavior of the connected correlation
function $\bar C(r) = \overline{\langle n_i \rangle\langle n_{i+r} \rangle -
\langle n_i n_{i+r} \rangle}$ [see Figs.~\ref{fig:correlations}(a-b)], which
decays algebraically as $\sim r^{-2}$ in the critical phase, similarly as for
the clean system~\cite{Alberton2021}. Deep within the area law, the correlations
decay more rapidly, as expected, while for the exceptional point $\gamma=0$,
exponential decay is present.

Interestingly, for $W = 0.5$ all collapses yield $\gamma_c$ estimates which are
higher than those for both $W = 0$ and $W = 1$; this implies that the
non-monotonicity of the phase boundary near small $W$ is a physical phenomenon,
and that introduction of weak disorder shifts the transition point $\gamma_c$ to
higher values. This behavior seems similar to the one observed in interacting
models~\cite{Boorman2021}, where a small amount of noise facilitated
entanglement spreading and extended the volume law. Here, a small amount of
disorder stabilizes the logarithmic phase, as the observed values of $c(L)$
appear to be higher [see Fig.~\ref{fig:limit_behavior}(a)]. Alternatively, a
weak disordered field slightly impedes the ability of measurements to pin the
wave function trajectory to the eigenstates of the measurement operator, so that
the area law occurs at higher $\gamma_c$.

\begin{figure}[tb]
	\centering
	\includegraphics[width = 0.8\columnwidth]{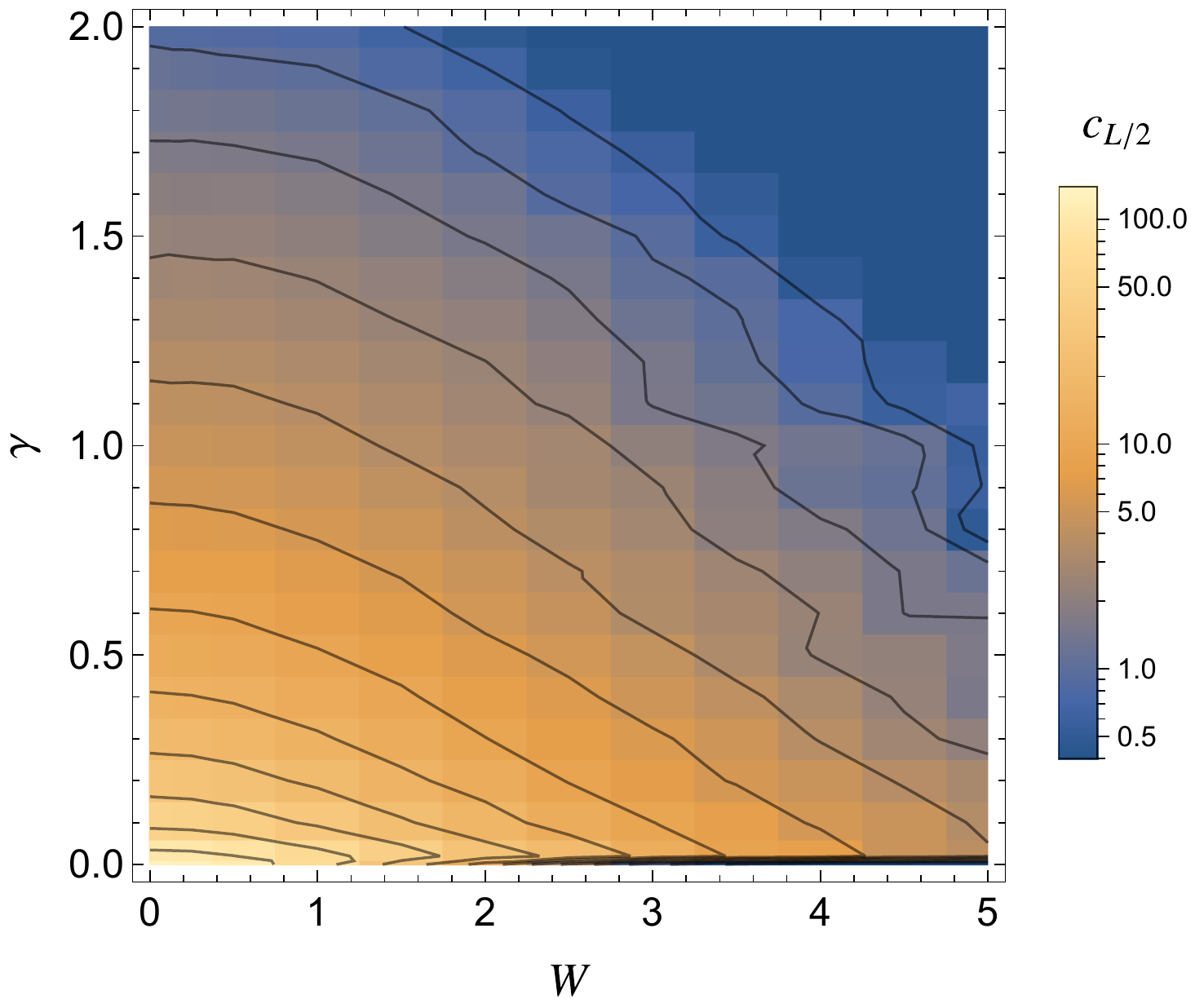}
	\caption{\changes{Central charge as a function of measurement strength $\gamma$ and 
	disorder $W$ for a system with additional next-nearest neighbor interactions.}}
  \label{fig:phase_diagram_nnn}
\end{figure}

\changes{Since the entanglement transition is a direct result of the competition between the unitary evolution and measurements, we believe that this non-monotonic behavior is tightly connected to the speed of entanglement spreading dictated by the hopping term. For this model, this speed seems low enough that the introduction of a small amount of disorder scrambles the information more efficiently, pushing $\gamma_c$ to higher values. We test this hypothesis by adding next-nearest neighbor interactions to the Hamiltonian,
\begin{equation}
  H = \sum_i ( c^\dagger_i c_{i+1} + c^\dagger_i c_{i+2} + \text{h.c.} )
    + \sum_i h_i n_i,
\end{equation}
which should increase the entanglement speed induced by the hopping terms. We find that the non-monotonicity in the phase diagram is absent (see Fig.~\ref{fig:phase_diagram_nnn}), as expected. This suggests that the n.n.n.\@ interactions increase the entanglement speed enough, so that it is no longer impacted by a small disorder.}

\changes{We would also like to comment on the recent results of Ref.~\cite{Poboiko2023}, which put into question the existence of the critical phase in the clean model. The non-monotonic behavior observed in the disordered free-fermion system studied here implies that the disorder may stabilize the critical phase, even if it is absent in the clean case, signifying the presence of the measurement-induced transition for a finite disorder strength.}

\subsection{Destruction of Anderson localization}

We now discuss the topic of small measurement strengths. For $\gamma = 0$, the
system becomes an Anderson insulator and exhibits area law for any finite $W$.
Below $W \lesssim 1.1$, finite size effects cause finite $c_{L/2}$: localization
length $\xi$ in the Anderson model is inversely proportional to
$W^2$~\cite{Kappus1981}, and $\xi$ becomes comparable to the considered system
sizes when $c_{L/2}$ becomes non-zero. This should, however, not be an issue for
larger $\gamma$, as the characteristic length $\xi$ is affected by both the
disorder and the measurements.

At very small but nonzero values of $\gamma = 0.02, 0.04$~\cite{Note1}, we
observe an abrupt change to the localized behavior. $\bar S(L/2)$ results
suggest that a logarithmic dependence on the system size is present for small
$W$ [Fig.~\ref{fig:small_param}(b)], with the crossover to the area-law
scaling at around $1.5 \lesssim W \lesssim 2.5$. We pinpoint the transition,
extracting $W_c^{\bar S}\approx 2.1$ and $W_c^{c(L)} \approx 1.9$ for
$\gamma=0.02$. Importantly, $W_c$ is large enough not to be impacted by the
characteristic length $\xi$ being comparable to $L$, and therefore we believe
the observed transition to be physical. Our results suggest the BKT universality
class is preserved for the whole transition boundary in
Fig.~\ref{fig:phase_diagram}(b). Furthermore, the connected correlators change
their behavior between $\gamma = 0$ and $0.02$ [see
Fig.~\ref{fig:correlations}(c-d)] from faster-than-algebraic to $1/r^2$ decay
for $W < W_c$, indicating emergence of the conformal phase.

\begin{figure}[tb]
  \centering
  \includegraphics[width = 0.99\columnwidth]{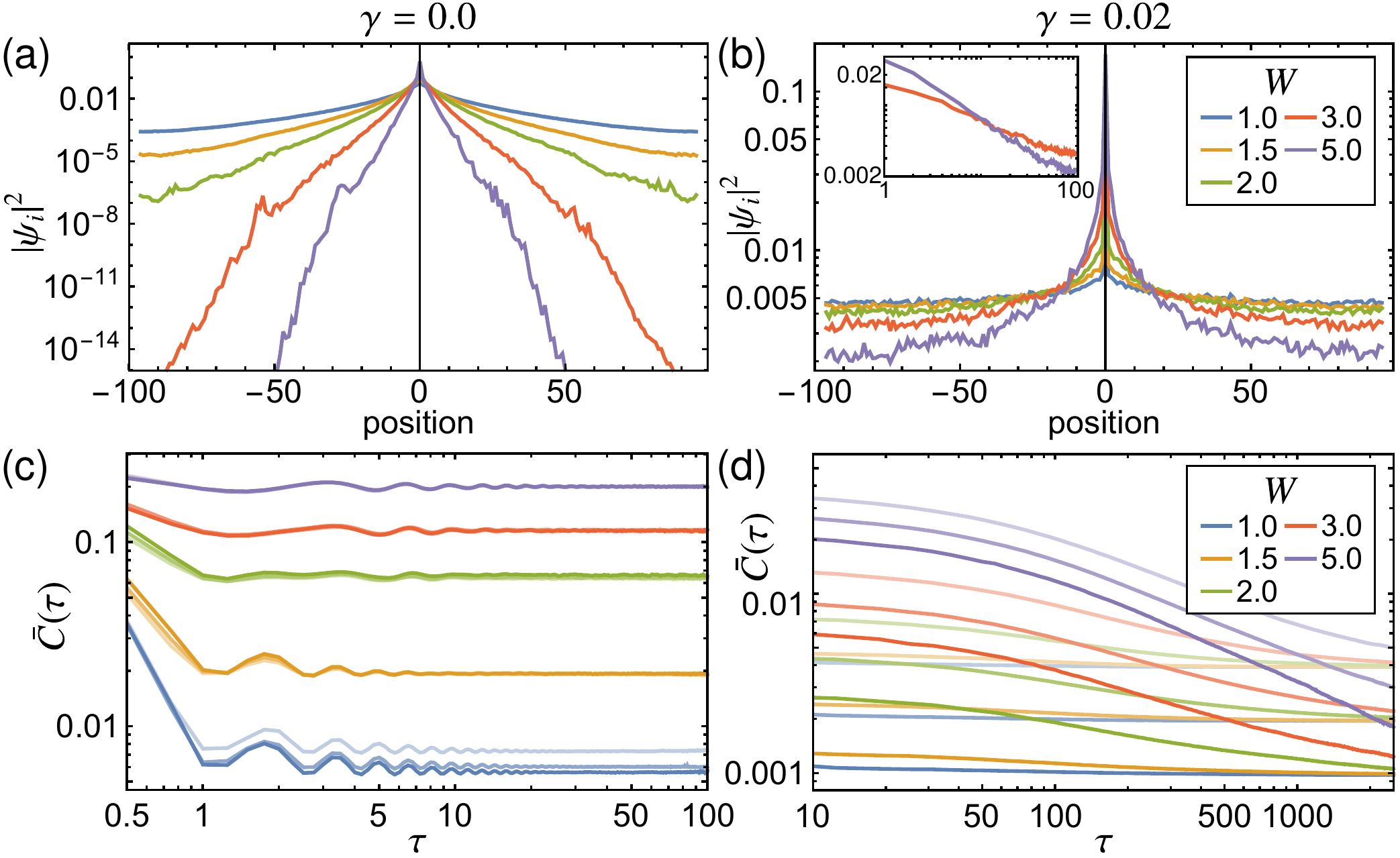}
  
  \caption{(a,b)~Averaged steady-state fermion orbitals $|\psi_i|^2$ for $L=192$
  for different values of $W$. Inset in (b) shows the log-log plot, suggesting
  power-law decay. (c,d)~Autocorrelation functions $\bar C(\tau)$ for different
  values of $W$ and system sizes $L=128$ (lighter), 256, 512 (darker). Left
  plots: $\gamma = 0$; right plots: $\gamma = 0.02$.} \label{fig:orbitals_auto}
\end{figure}

We thus conclude that Anderson localization is immediately broken for any finite
value of $\gamma$, and the critical phase reappears in the phase diagram. This
phenomenon most likely occurs due to measurements impeding interference through
impurity scattering---even very weak measurements change the scattered fermionic
modes and destructive interference is not possible, rendering the mechanism
behind Anderson localization disrupted. \changes{A similar mechanism occurs when inelastic scattering is introduced to an Anderson-localized medium, where the phase coherence between outgoing and ingoing modes is disrupted~\cite{Altshuler1998}. Whether the measurements force the
system into the critical phase or the area law depends on the shape of
localized single-particle orbitals $|\psi_i|^2$ at $\gamma = 0$ [see
Fig.~\ref{fig:orbitals_auto}(a,b)], which can be read off from the Slater determinant. At large disorder, the orbitals decay rapidly and the overlap between their envelopes is
negligible. The measurements have little impact, only sharpening the orbitals at their localization centers, and the area law is preserved. At small disorder, the orbitals are broad
and their envelopes substantially overlap with each other. The measurements effectively introduce
scrambling between them, which leads to a delocalized behavior and the critical
phase. This very simple picture would suggest that the transition happens approximately when $\xi \sim 1$, while our numerical results reveal a slightly larger critical localization length of $\xi \sim 24/W_c^2 \approx 6$~\cite{Kappus1981}.}

Furthermore, we find a clear distinction between the Anderson-localized area law
and the measurement-induced area law for $\gamma > 0$. The former is
characterized by \textit{exponential decay} of the orbitals, while the latter
exhibits \textit{power-law localization}, $|\psi_i(x)|^2 \sim x^{-\alpha}$
[inset in Fig.~\ref{fig:orbitals_auto}(b)]~\cite{Note1}. Autocorrelation
functions $\bar C(\tau)$ [Fig.~\ref{fig:orbitals_auto}(c,d)] also showcase this
difference. For $\gamma = 0$, $\bar C(\tau)$ quickly saturates to a constant and
does not decay. However, for $\gamma > 0$, $\bar C(\tau)$ plateaus for a long
period of time ($\tau \sim 100$), and eventually decays due to the disruption
from measurements \changes{to a minimum value of $1/(2L)$}. The plateau size depends on $\gamma$, where for large
$\gamma$ the decay begins earlier. The decay itself seems to be approximately
power law, with larger systems taking more time to reach the minimum value.

\section{Conclusions}

The results presented in this \changes{work} show the nontrivial interplay between
Anderson localization and continuous measurements. We convincingly demonstrate
that the entanglement phase transition from the critical phase with conformal
symmetry to the area-law phase survives the introduction of quenched disorder.
Moreover, the universality class of this transition also seems to be preserved,
which strongly suggests that the logarithmic phase is stable to weak
perturbations. We also find that a small amount of disorder can help stabilizing
the critical phase. Gathering all our data from the collapse of entanglement
entropy and effective central charge, we estimate the true transition boundary
between the logarithmic and area-law phases [solid line in
Fig.~\ref{fig:phase_diagram}(b)]. In general, our results convincingly suggest
the conformal phase and free-fermion MIETs are viable for experimental probing
in systems such as nanowires and quantum dot arrays, which host Anderson
localization along with implementation of local measurements~\cite{Kim_PRX2022}.

We find that an introduction of monitoring in the Anderson-localized model
results in an instant destruction of the localization for weak disorder. The
delocalization results from the destruction of the coherent processes leading to
a liquid state. Although, at sufficiently large disorder, the system transitions
into an area-law state which is markedly distinct from Anderson localization as
the orbitals exhibit a power-law decay in space instead of exponential. The
temporal behavior of the autocorrelations exhibits parametrically longer decay
time scales compared to Anderson localization. There are several interesting
directions for future work emerging from our results. The role of interactions
in the logarithmic phase and its relationship to many-body localization remains
a challenging open problem. The fate of the critical phase for integrable models
which do not map to free fermions could also provide new classes of
measurement-induced criticality.

All relevant data present in this publication can be accessed at~\cite{data}.

\begin{acknowledgments}
  We thank Romain Vasseur, Sebastian Diehl, Alessandro Romito, Paul P\"opperl, Igor Poboiko, and Igor Gornyi for useful discussions.
  A.P.\ and M.S.\ were funded by the European Research Council (ERC) under the European Union's Horizon 2020 research and innovation programme (grant agreement No.\ 853368).
  O.L.\ was supported by UKRI grant MR/T040947/1.
  The authors acknowledge the use of the UCL Myriad High Performance Computing Facility (Myriad@UCL), and associated support services, in the completion of this work.
\end{acknowledgments}







\appendix

\section{Methodology}
\label{app:method}

The evolution considered in this paper preserves particle number and can be
efficiently simulated with a method based on stochastic Schr{\"o}dinger
equation~\cite{Cao2019}. The wave function is a pure Gaussian state of $N$
particles on $L$ sites and can be described by an $N \times L$ matrix $U$,
\begin{equation}
  | \psi \rangle = \prod_{k = 1}^N \left( \sum_{j = 1}^L U_{j k} c_j^{\dagger}
  \right) | 0 \rangle,
\end{equation}
where $c_j^{\dag}$ are the fermionic creation operators, and $| 0 \rangle$ is
the vacuum state. \changes{Physically, $U$ is a matrix of fermion orbitals (single-particle wave functions), and $\det(U)$ is a Slater determinant.} In this work,
we always consider a case of half-filling and start the evolution from a Ne\'el
state.

The measurements and time evolution are implemented using the stochastic
Schr\"{o}dinger equation, where a monitoring of an operator $O$ is done by
evolving the wave function according to
\begin{equation}
  d | \psi (t) \rangle = - i H d t | \psi (t) \rangle +\mathcal{M} | \psi (t)
  \rangle,
\end{equation}
where measurement operator is $\mathcal{M}= \left( (O - \langle O
\rangle_t) d \eta_t - \frac{\gamma}{2} (O - \langle O \rangle_t)^2 d t \right)$,
with $\eta_t$ a Wiener process and $\gamma$ the measurement strength/rate. We
will measure operator $n_i = c_i^{\dag} c_i$ on every site. This
evolution can be approximated by trotterisation, $| \psi (t + d t) \rangle
\approx e^{\mathcal{M}} e^{- i H d t} | \psi (t) \rangle$.

Importantly, this corresponds to an evolution of the matrix $U$ that fully
describes the Gaussian state,
\begin{equation}
  U (t + d t) = e^M e^{- i h d \nospace t} U (t),
\end{equation}
where $M$ is a matrix with elements $M_{i j} = \delta_{i j} (\eta_i + (2 \langle
n_i \rangle -1) \gamma d t)$, and $h$ corresponds to the free-fermion
Hamiltonian $H = \sum_i c_i^{\dag} c_{i + 1} + h.c.$, and has elements $h_{i j}
= \delta_{i, j + 1} + \delta_{i, j - 1} + h_i\delta_{i, j}$. After each time
step $d t$, the wave function needs to be properly normalized, which can be done
by a QR decomposition of matrix $U (t + d t) = Q R$, and setting the new matrix
$U$ to be $Q$.

\begin{figure}[tb]
	\centering
	\includegraphics[width = 0.99\columnwidth]{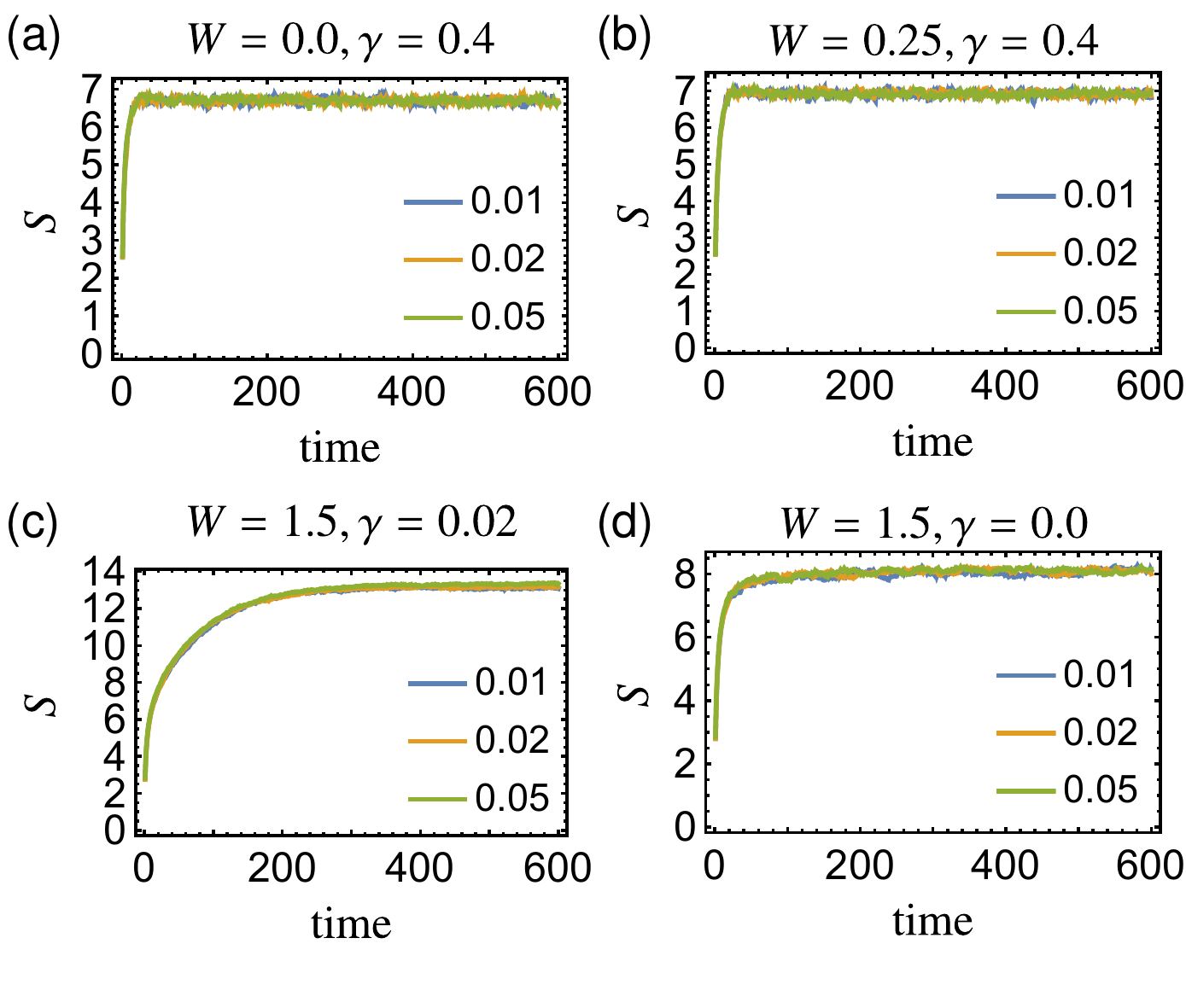}
	\caption{Averaged entanglement entropy $S$ as a function of time for four values of disorder $W$ and measurement strength $\gamma$: (a) $W = 0.0, \gamma = 0.4$, (b) $W = 0.25, \gamma = 0.4$, (c) $W = 1.5, \gamma = 0.02$, and (d) $W = 1.5, \gamma = 0.0$. The system size is $L = 256$. Different colors signify the time steps $dt = 0.01, 0.02, 0.05$. The curves collapse well for all used time steps. At long times, entropy reaches the saturation value $S_\infty$.} \label{fig:time_collapse}
\end{figure}

Fig.~\ref{fig:time_collapse} shows that setting $d t = 0.05$ is enough to
describe the continuous-time regime, and we find that lowering $dt$ does not
change our results within the statistical error bars.

\subsection{Observables}

\begin{figure}[tb]
	\centering
	\includegraphics[width = 0.85\columnwidth]{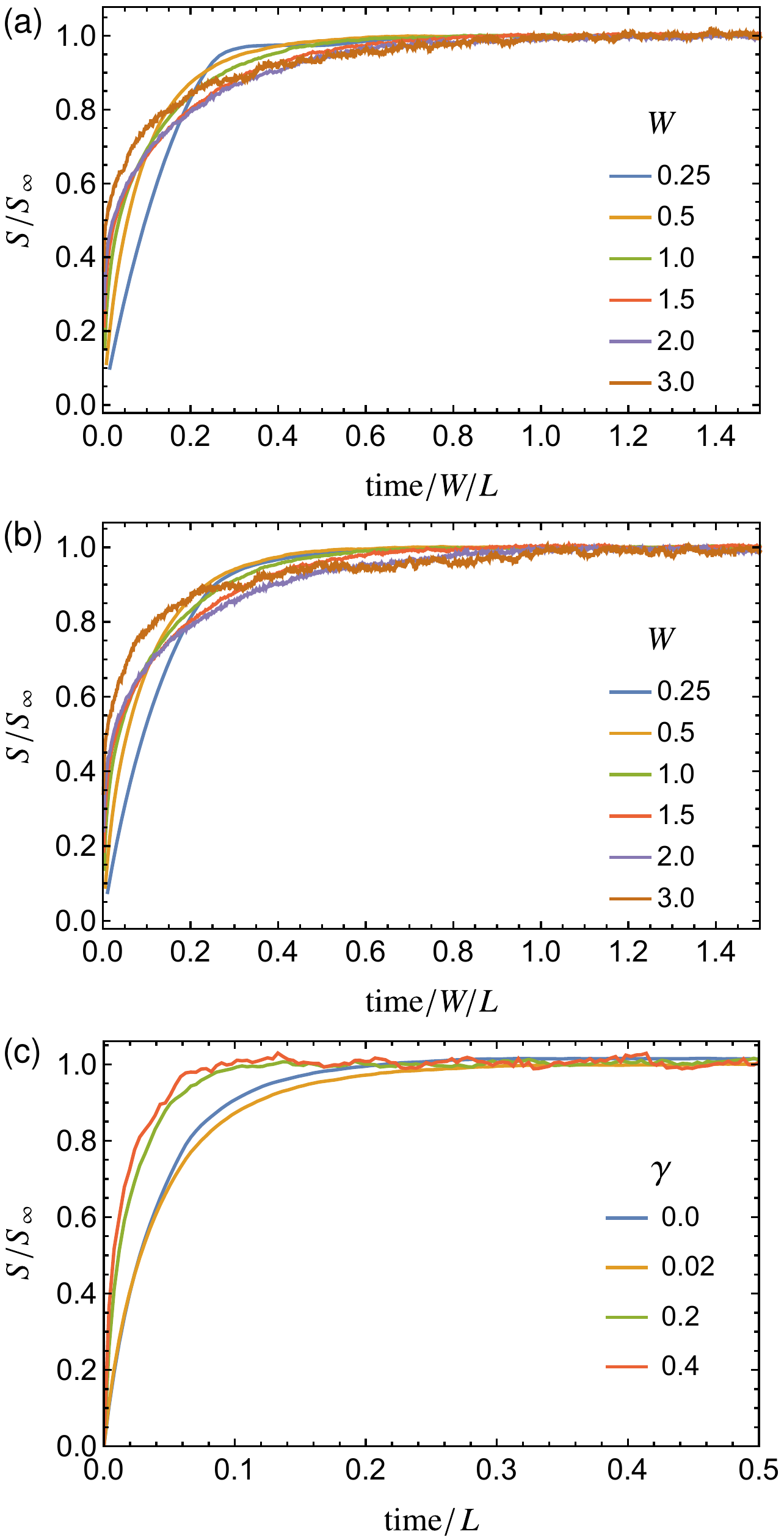}
	\caption{\changes{Time dependence of averaged entanglement entropy $S$ for (a)~$\gamma = 0.02, L = 256$, (b)~$\gamma = 0.02, L = 385$, and (c)~$W = 0.5, L = 256$. The time after which the value reaches saturation value is roughly proportional to the system size $L$, but is also impacted by the disorder strength $W$ and the measurement rate $\gamma$.}}
  \label{fig:equil}
\end{figure}

Using the matrix $U$, one can define the correlation matrix $D = U
U^{\dagger}$ with elements $D_{i j} = \langle c_i^{\dagger} c_j \rangle$,
giving us direct access to expectation values. Furthermore, to calculate
entanglement entropy $S$ of a bipartition of the system into subsystem A and
its compliment B, we restrict $D$ to indices associated with the subsystem A,
and then diagonalize the restricted matrix to obtain its eigenvalues
$\lambda_i$. $S$ is then simply given by
\begin{equation}
  S = - \sum_i \left( \lambda_i \ln \lambda_i + \vphantom{\frac{}{}} (1 -
  \lambda_i) \ln (1 - \lambda_i) \right) .
\end{equation}

The connected correlation functions $C(r)$ can be determined from the correlation matrix,
\begin{equation}
  C(r) = | D_{i+r,i} |^2 = \langle n_i \rangle \langle n_{i+r} \rangle - \langle n_i n_{i+r} \rangle.
\end{equation}
Similarly, the autocorrelation function $C(\tau)$ can be calculated in the same manner,
\begin{equation}
  C(\tau) = | D_{i,i}(t,t+\tau) |^2,
\end{equation}
where $D(t,t+\tau) = U(t+\tau) U^\dagger(t)$.

\begin{figure}[t]
  \centering
  \includegraphics[width = 0.8\columnwidth]{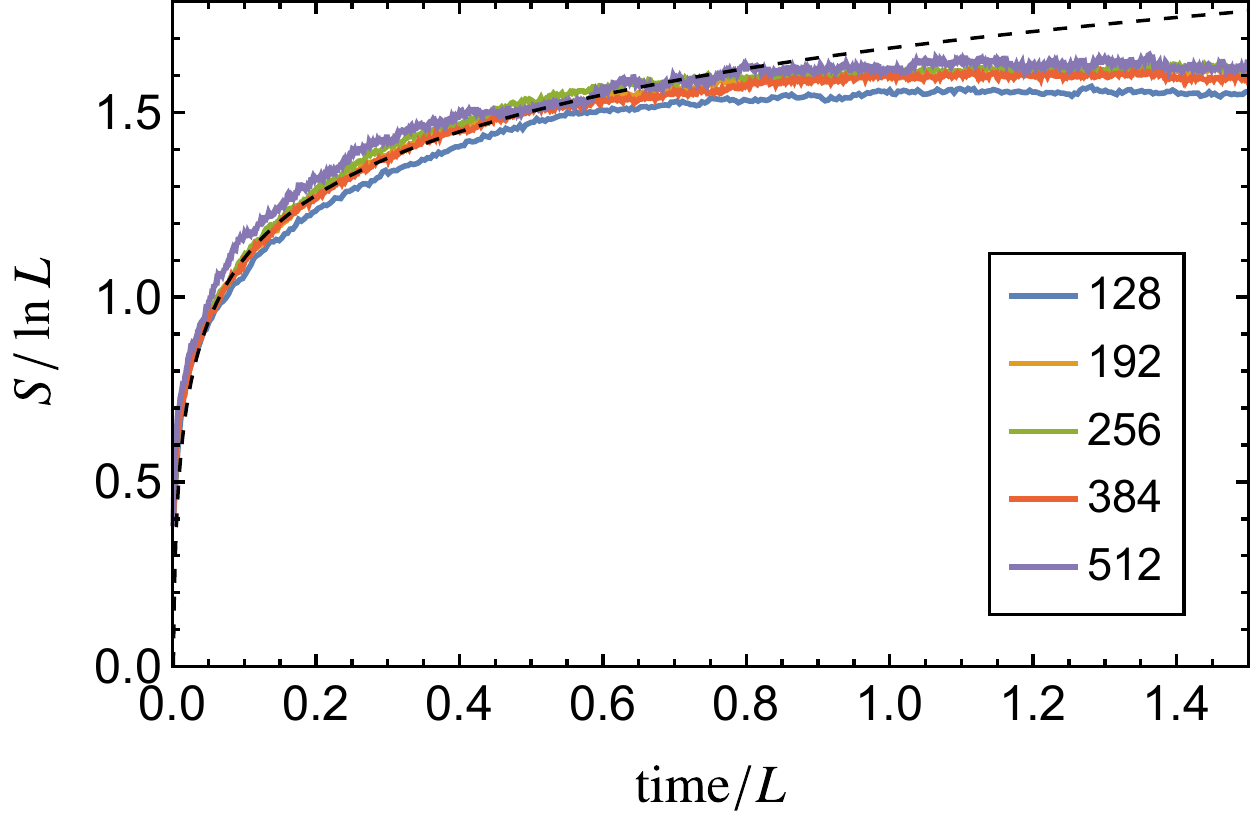}
  
  \caption{Emerging conformal symmetry near $W=2$ for $\gamma=0.02$.
  Entanglement entropy is rescaled as $S/\ln L$, while the time is scaled as
  $\text{time}/L$. Dashed line is a fit to a logarithmic behavior.}
  \label{fig:conf_sym}
\end{figure}

\begin{table}[t]
  \begin{tabular}{lccccc}
    \hline \hline
    & Entropy & \, & \multicolumn{3}{c}{Central charge} \\
    \cline{2-2}  \cline{4-6}
    Data & $\gamma_c^{\bar S}$ or $W_c^{\bar S}$ && $\gamma_c^{c(L)}$ or 
    $W_c^{c(L)}$ & $\alpha$ & $\beta$\\
    \hline
    $W = 0.0$~\cite{Alberton2021} & 0.31(5)  && 0.21(5) & 3.99 & 4.37 \\
    $W = 0.25$                    & 0.35(5)  && 0.31(5) & 4.00 & 5.4  \\
    $W = 0.5$                     & 0.40(6)  && 0.35(5) & 4.1  & 7.6  \\
    $W = 1.0$                     & 0.33(5)  && 0.29(5) & 5.12 & 7.65 \\
    $\gamma = 0.02$               & 2.06(15) && 1.92(25)& 6.4  & 7.8  \\
    $\gamma = 0.04$               & 2.07(15) && --      & --   & --   \\
    \hline \hline
  \end{tabular}
  \caption{Data collapse parameters for the entropy and central charge results.}
  \label{tab:collapse_params}
\end{table}

Finally, one can easily extract the fermion orbitals by taking the columns of
$U$, i.e. $|\psi_i(r)|^2 = |U_{i,r}|^2$. We move the orbitals spatially so that
they are centered around the maximum value, and then average them over many
realizations.

\subsection{Equilibration to the steady state}

The time it takes to reach the steady state is nontrivially dependent on two
variables: measurement strength $\gamma$ and disorder $W$. In the absence of the
disorder, for large $\gamma$ we find that the equilibration takes $\mathcal{O}(1)$ time, while
for small $\gamma$ it takes at most $\mathcal{O}(L)$ time. Introducing the disorder prolongs
the equilibration time roughly proportionally to $W$ (see Fig.~\ref{fig:equil}).

We also note that near $W \approx 2$, the time dependence of the
trajectory-averaged half-chain entropy seems to be collapsing into one curve
[see Fig.~\ref{fig:conf_sym}], with the initial behavior scaling as $S(L/2)\sim
\ln (\text{time}/L)$. This suggests an emergence of $z=1$ conformal symmetry
near this point.

\subsection{Finite size scaling}

\begin{figure}[t]
  \centering \includegraphics[width = 0.99\columnwidth]{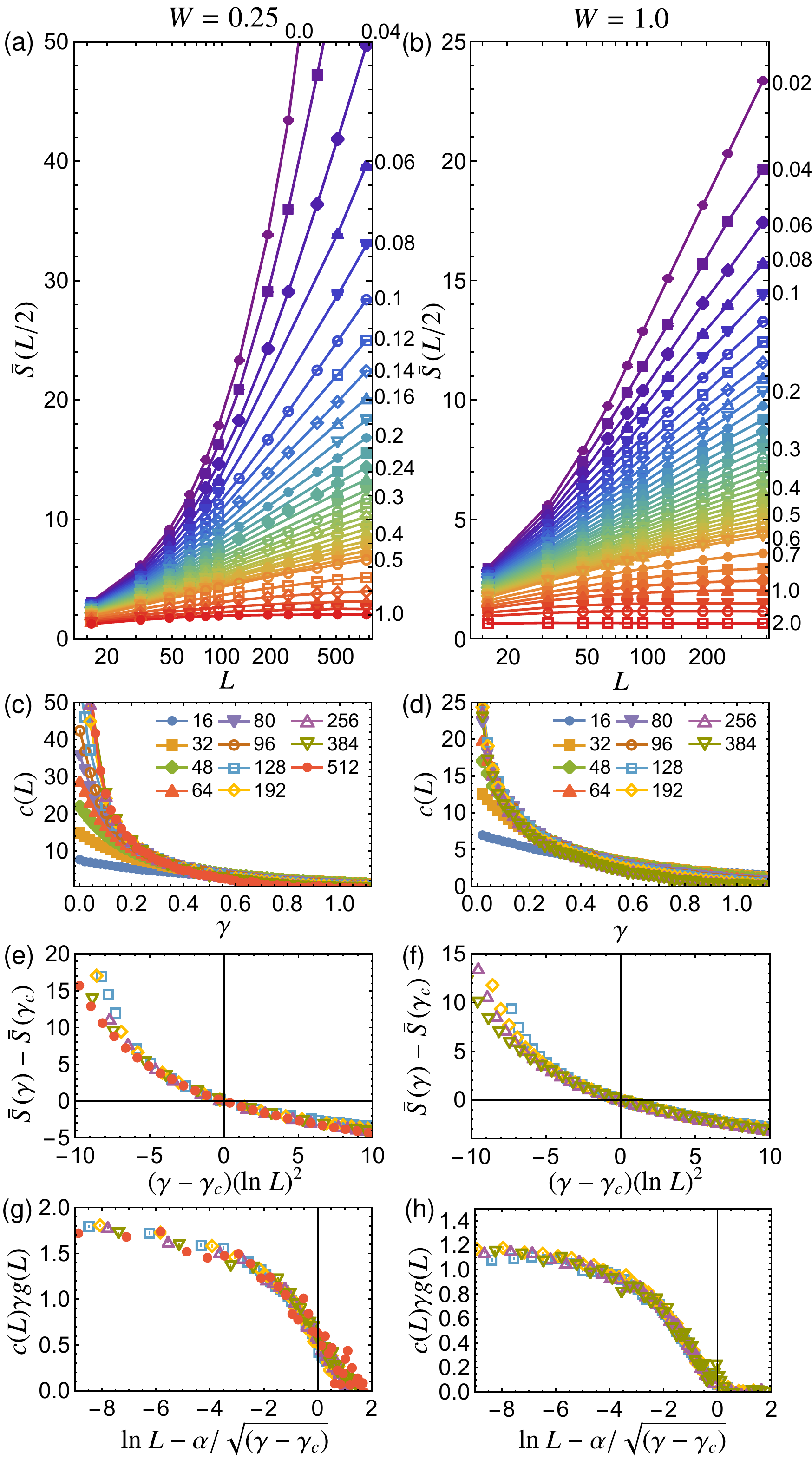}
  
  \caption{Behavior of (a,b)~half-chain entanglement entropy $\bar S(L/2)$ for
  different values of the measurement strength $\gamma$ (see labels on the
  right), and (c,d)~central charge $c(L)$. Data collapse for (e,f) $S(L/2)$, and
  (g,h) $c(L)$; legend from (c,d) applies in (e-h). Left plots are for $W =
  0.25$ and the right plots are for $W = 1.0$.}
  \label{fig:supporting_data}
\end{figure}

\begin{figure}[t]
  \centering \includegraphics[width = 0.99\columnwidth]{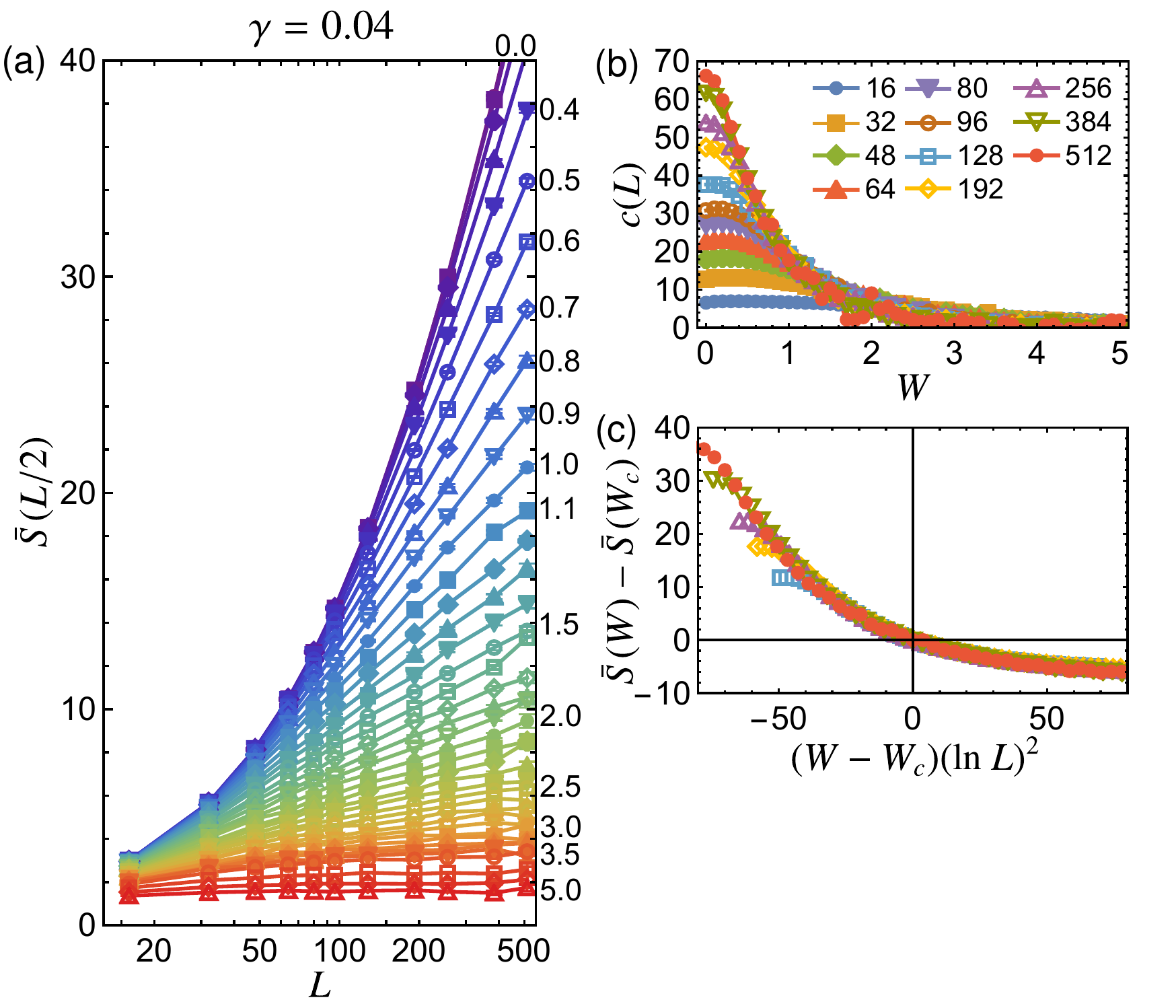}
  
  \caption{Results for $\gamma=0.04$. Behavior of (a)~half-chain entanglement entropy $\bar S(L/2)$ for different values of $W$ (see
  labels on the right), and (b)~central charge $c(L)$. Data collapse for
  (c)~$S(L/2)$; legend from (b) applies in (c).} \label{fig:supporting_data2}
\end{figure}

The data collapse for the finite-size scaling analysis is performed by
minimizing the cost function $\epsilon$, which measures how well the data
collapses into a single curve given the parameters. First, the data is rescaled
using the finite-size scaling ansatze from Eqs.~(\ref{eq:S_data_collapse}) and
(\ref{eq:c_data_collapse}) to produce a set of triples ${x_i, y_i, d_i}$
representing the rescaled x coordinate, rescaled y coordinate, and the error in
the y coordinate. For example, for Eq.~(\ref{eq:S_data_collapse}), $x = (\gamma
- \gamma_c^{\bar S}) (\ln L)^2$, $y = \bar S(L/2,L,\gamma) - \bar
S(L/2,L,\gamma_c^{\bar S})$, and $d$ is the error of the half-chain entropy.
Then, the triples are sorted by their x-values, and one can calculate the cost
function,
\begin{equation}
  \epsilon = \frac{1}{n{-}2} \sum_{i=2}^{n{-}1} w(x_i, y_i, d_i | x_{i{-}1}, y_{i{-}1}, d_{i{-}1}, x_{i{+}1}, y_{i{+}1}, d_{i{+}1}),
\end{equation}
where
\begin{align}
  w & = \frac{(y_i - \bar y)^2}{\Delta^2}, \\
  \bar y & = \frac{(x_{i+1} - x_i)y_{i-1} - (x_{i-1}-x_i)y_{i+1}}
    {x_{i+1}-x_{i-1}}, \\
  \Delta^2 & = d_i^2
    + \frac{(x_{i+1}-x_i)^2 d^2_{i-1} + (x_{i-1}-x_i)^2 d^2_{i+1}}
    {(x_{i+1}-x_{i-1})^2}.
\end{align}
After obtaining the minimum $\epsilon_\text{min}$, one can estimate the error in
the collapse parameters by investigating the region where $\epsilon = 2
\epsilon_\text{min}$.

In Table~\ref{tab:collapse_params} we report the estimates for the parameters
from data collapses in Figs.~\ref{fig:small_param}, \ref{fig:supporting_data}, and \ref{fig:supporting_data2}. \changes{The scaling function $g(L)$ from Eq.~(\ref{eq:c_data_collapse}) has the following form, $g(L) = [1+1/(2\ln L - \beta)]^{-1}$, and can be determined from a superfluid stiffness scaling analogy for the BKT transition~\cite{Harada1997, Carrasquilla2012, Alberton2021}.}

\begin{figure}[t]
  \centering
  \includegraphics[width = \columnwidth]{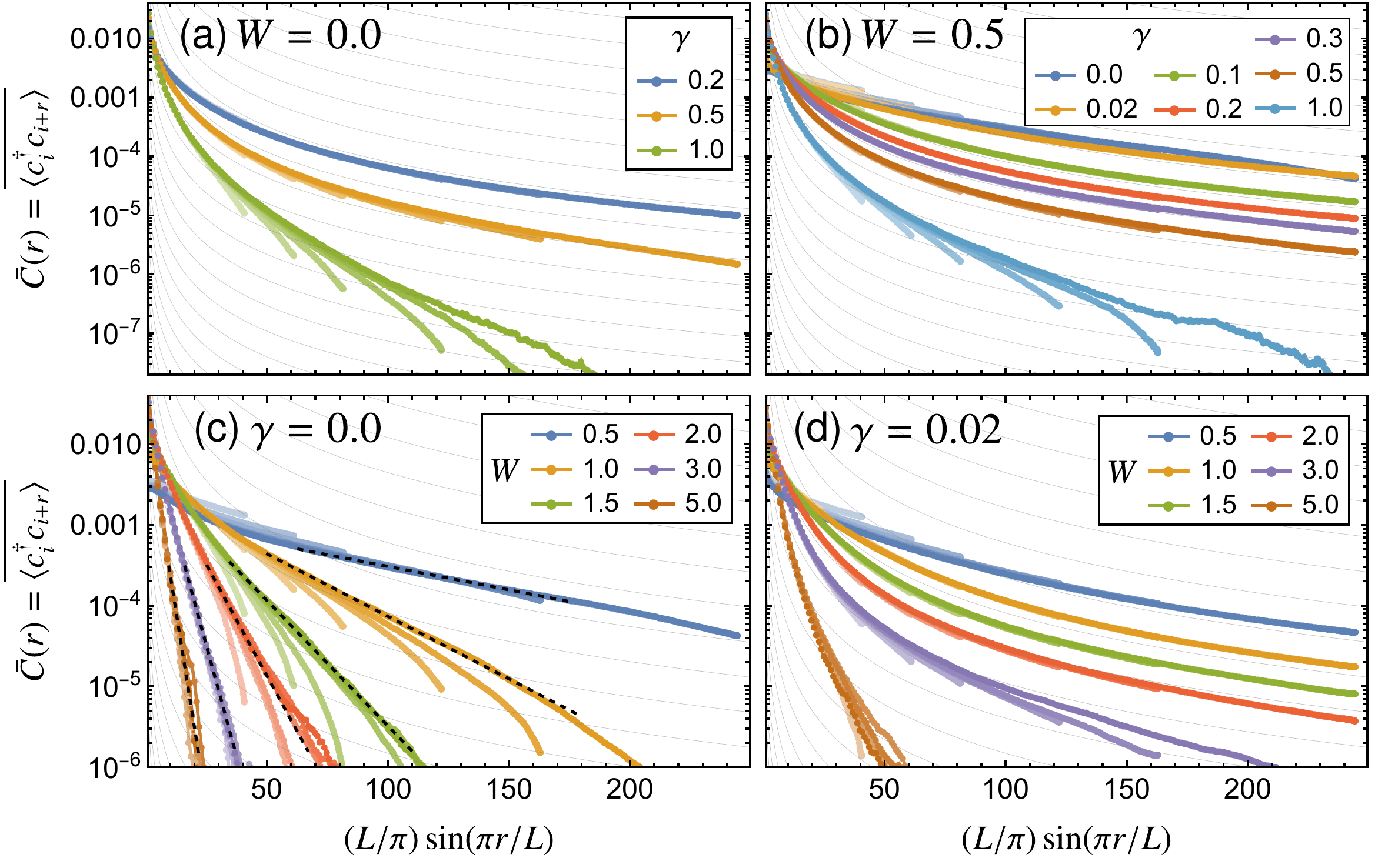}
  
  \caption{Connected correlation function $\bar C(r)$ for constant disorder strength (a) $W=0.0$ (no disorder), (b) $W=0.5$, and constant measurement strength (c) $\gamma=0.0$, (d) $\gamma=0.02$. Plot opacity indicates the system size ($L = 128, 196, 256, 384, 512, 768$). Gray lines show the algebraic decay of $\sim r^{-2}$ expected for the critical phase. Dashed lines in subfigure (c) show exponential decay for Anderson localization.} \label{fig:correlations2}
\end{figure}

Supporting data for $W = 0.25$ and $\gamma = 0.04$ is shown in 
Fig.~\ref{fig:supporting_data}.

\subsection{Decay of correlation functions}

Although we find that single-particle wave functions are power-law localized, the issue is that these orbitals are not uniquely defined, as the matrix $U$ can be multiplied on the right by any unitary, while not changing the physical state. However, we also find that for $\gamma>0$, correlation functions do not seem to exhibit exponential decay (see Fig.~\ref{fig:correlations2}, where we show the data of Fig.~\ref{fig:correlations}, but on a linear-log plot), which would be in agreement with our findings for the orbitals. Perhaps the reason why we do not find ``scrambled'' orbitals, is due to the uniqueness of the method: both unitary evolution and measurements uniquely transform matrix $U$ (during normalization, QR decomposition is unique as well).

\clearpage

\bibliography{refs}

\end{document}